\documentstyle[prl,twocolumn,aps]{revtex}
\input{epsf}
\begin{document}
\draft
\twocolumn[\hsize\textwidth\columnwidth\hsize\csname @twocolumnfalse\endcsname
\title{Emergence of quantum chaos in finite interacting Fermi systems}

\author{Ph. Jacquod$^{(a)}$ and D. L. Shepelyansky$^{(b, *)}$}

\address {$^{(a)}$ Institut de Physique, Universit\'e de Neuch\^atel,
1, Rue A.L. Breguet, CH-2000 Neuch\^atel, Suisse \\
$^{(b)}$ Laboratoire de Physique Quantique, UMR C5626 du CNRS, 
Universit\'e Paul Sabatier, F-31062 Toulouse Cedex 4, France}

\date{5 June, 1997}

\maketitle

\begin{abstract}
We study the level spacing statistics $P(s)$ 
in many-body Fermi systems and determine 
a critical two-body interaction strength $U_c$
at which a crossover from Poisson to Wigner-Dyson
statistics takes place. Near the Fermi level the results 
allow to find a critical temperature $T_{ch}$ above which
quantum chaos and thermalization set in.
\end{abstract}
\vskip1pc]

\narrowtext


The Random Matrix Theory (RMT) was developed to explain the general properties
of complex energy spectra in many-body interacting systems such as heavy nuclei, 
many electron atoms and molecules \cite{rmt}. Later, it has found many other 
successful applications in different physical systems. Among the most recent of
them we can quote models of quantum chaos where RMT appears due
to the classically chaotic but deterministic underlying dynamics \cite{bohigas}.
One of the most direct indications of the emergence of quantum chaos is the
transition of the level spacing statistics $P(s)$ from Poisson to 
Wigner-Dyson (WD) distribution. This property has been widely used
to detect the transition from integrability to chaos not only in systems with 
few degrees of freedom \cite{bohigas} but also in solid-state models with many
interacting electrons \cite{bel}. It was also applied to determine the  
delocalization threshold in noninteracting disordered systems \cite{shklo}.

While the conditions for the appearance of the WD distribution in 
noninteracting systems is qualitatively well understood the situation is more
intricate in presence of interaction. Indeed, in this case the size of
the total Hamiltonian matrix grows exponentially with the number of 
particles and it becomes
very sparse as a result of the two-body nature of the interaction. 
Due to that it was initially not obvious whether switching on the interaction 
would lead to the WD statistics. 
To study this problem a two-body random interaction
model (TBRIM) had been proposed \cite{french,bohigas1}. 
This model consists of $n$ fermions which can occupy $m$ unperturbed energy 
orbitals with mean one-particle level spacing $\Delta$. 
The multiparticle states are coupled by two-body random transition
matrix elements of typical strength $U$. It was found that a
sufficiently strong $U$ leads to a level mixing and appearance of
WD statistics. Very recently the interest for this model has been renewed
and its statistical properties were investigated in more details \cite{flam}.
This raise of interest was stimulated by the understanding that many
statistical properties of real physical systems such as 
the rare-earth Ce atom \cite{flam1}
and the $^{28}$Si nucleus \cite{zel,zel1} are well described by the TBRIM. In
addition this model is quite similar to the {\it s-d} shell model used for a
description of complex nuclei \cite{zel,zel1}. Since interaction is generically
of two-body nature it is reasonable to assume that this model will be 
also useful
for a description of interacting electrons in clusters \cite{akulin} and 
mesoscopic quantum dots \cite{sivan}.

While the statistical properties of the TBRIM were studied in some details,
surprisingly, the most important question of the critical interaction strength 
$U_c$ at which the WD level spacing statistics sets in was omitted. 
Apparently the
reason for this is based on the common lore in nuclear physics that the level 
density grows exponentially with the number of particles and therefore an
exponentially small interaction is sufficient to mix nearby levels
\cite{flam,zel1}. However recent estimates on few-particle models
($n=2,3,4$) showed that in spite of the high many-body density of states,
only an interaction strength comparable to the two-particle level spacings 
can give a level mixing \cite{oleg,pich}. Therefore the dependence of $U_c$ on 
the number of particles and orbitals as well as the excitation energy should
still be determined. This is the main purpose of this paper. 
The above border in $U$
is physically very important. Indeed for $U<U_c$ levels are not mixed by
interaction and hence the system is not thermalized. Consequently the 
occupation numbers are
not described by the Fermi-Dirac statistics. On the contrary a
sufficiently strong interaction leads to thermalization as it has been seen
in numerical simulations \cite{flam,zel,zel1}.

To study the effect of interaction on the spectral properties of finite
Fermi systems we used the TBRI model described in \cite{flam}. It consists
of $n$ particles distributed over $m$ orbitals with energies $\epsilon_{m'}$
, $m'=1,2,...m$. These energies are randomly distributed over the interval
$[0,m]$ with average spacing $\Delta=1$. The total number of multiparticle
states is $N=m!/(n!(m-n)!)$. They are coupled by random two-body transition
matrix elements distributed in the interval $[-U,U]$. Due to the two-body
nature of the interaction, only states differing by at most two one-particle
indices are coupled. As a result each multiparticle state is coupled with
$K=1+n(m-n)+n(n-1)(m-n)(m-n-1)/4$ states \cite{flam}. All these transitions
occur inside a two-body energy interval $B=2m-4$ around the energy of an
initial multiparticle state. For large $m$ and $n$, the number of transitions
$K$ is much smaller than the size of the matrix $N$ but  is much
larger than the number of different two-body matrix elements 
$N_2 \approx m^2/2$. The total energy of the system varies from the
ground-state value $E_{g} \approx n^2 \Delta/2 $ to the maximal value $E_t
\approx m n \Delta -E_{g}$ and the Fermi energy is $\epsilon_F \approx n \Delta$. The
typical level spacing in the middle of the spectrum at $E_h \approx (E_t+E_g)/2$
is $\Delta_n \approx (E_t-E_g)/N$.

Let us first discuss the situation at high energies $E \sim E_h$
where all $K$ transitions are energetically allowed. In this case the density
of {\it directly} coupled states is $\rho_c \approx K/B$ 
because all transitions take
place inside the two-body energy band $B$. According to perturbation theory
these levels will be mixed when the transition matrix element $U$ between 
them becomes of the order of the corresponding spacing $\Delta_c = 1/\rho_c$.
This determines the critical coupling $U_c$
\begin{equation}
\label{uc}
U_c = C \frac{B}{K} \approx \frac{2 C}{\rho_2 n^2}
\end{equation}
Here, we introduced the two-particle density $\rho_2 \approx N_2/B \approx m/4$
assuming $m \gg n \gg 1$ and a numerical constant $C$ to be determined. 
For $U \ll U_c$ the perturbation theory works, levels are not 
mixed and $P(s)$ is
close to the Poisson distribution. For $U>U_c$ we expect a strong mixing of
levels not only on a scale $\Delta_c$ but on a much smaller scale $\Delta_n$.
There are few arguments in favor of this statement. The first of them is based
on the results for few-particle systems ($n=2,3,4$) \cite{oleg}.
According to \cite{oleg}, the effective transition 
matrix element between nearby levels in high orders of perturbation theory 
becomes comparable to $\Delta_n$ when the first-order transition mixes directly
coupled states ($U > U_c$). Recently the same conclusion was 
drawn in \cite{pich}. The second argument is based on an analogy with
superimposed band
random matrices (SBRM) with strongly fluctuating diagonal elements
\cite{lenz,jac,yan,klaus}. There it was shown that for sufficiently large
band (number of nonzero diagonals $2b+1 \gg \sqrt{N}$) the eigenstates are
extended over the whole matrix size $N$ and $P(s)$ has the WD form if the
transition matrix elements are larger than the energy spacing between directly 
coupled states. This condition is rather similar to the above border
(\ref{uc}).

To check the prediction (\ref{uc}), we numerically
computed $P(s)$ in the middle of the spectrum of
the TBRIM (keeping only $\pm 25\%$ of the levels around $E_h$) 
for $n \leq 8$ and
$m \leq 80$ at various interaction strengths $U$.
Up to 5000 different realizations of disorder have been used to obtain
the total spacing statistics $N_s \approx 30000$.
A typical example of the transition from Poisson to WD statistics is shown in
Fig. 1. As expected the level repulsion disappears at small $U$ while for large
$U$ the distribution approaches the WD form. To characterize this transition we
computed for each distribution $P(s)$ the value $\eta = \int_0^{s_0}
(P(s)-P_{WD}(s)) ds / \int_0^{s_0} (P_{P}(s)-P_{WD}(s)) ds$. Here $P_{P}(s)$
and $P_{WD}(s)$ are the Poisson and the WD distributions respectively and
$s_0=0.4729...$ is their intersection point. In this way $\eta$ varies from 1
($P(s)=P_P(s)$) to 0 ($P(s)=P_{WD}(s)$). We determined the critical interaction
strength $U_c$ by the condition $\eta(U_c) = \eta_c = 0.3$. The choice of
$\eta_c$ influences only the numerical factor $C$ in (\ref{uc}). We note that
this $\eta$-value is close to the value $\eta_A = 0.215$ corresponding to
$P(s)$ at the Anderson transition in 3d \cite{shklo} (in \cite{shklo} a
criterion slightly different from ours was used).

\begin{figure}
\epsfxsize=3.7in
\epsfysize=2.6in
\epsffile{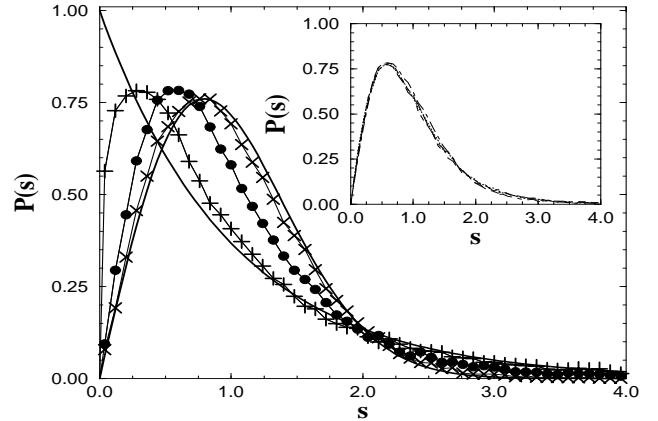}
\vglue 0.2cm
\caption{Transition from Poisson to Wigner-Dyson statistics in the TBRIM for
$m=12$, $n=6$ : $U/\Delta=0.01$ and $\eta=0.93$ (+); 
$U/\Delta=0.055$ and $\eta=0.3$
($\bullet$); $U/\Delta=0.13$ and $\eta=0.063$ (x). 
Full lines show the Poisson and the
Wigner-Dyson distributions. Insert shows $P(s)$ at fixed
$\eta=0.3$ for half-filling $\nu=n/m=0.5$ and $n=4$ (dotted line), $n=5$ 
(dashed line), $n=6$ (long dashed line) and $n=7$ (dotted-dashed line).
} 
\label{fig1}
\end{figure}

The fact that the concrete choice of $\eta_c$ is not crucial is also confirmed 
by Fig. 2 which shows the existence of a scaling $\eta = \eta(U/U_c)$. Indeed
the numerical data in a large parameter range demonstrate the existence of
one scaling curve (Fig. 2). This scaling is very similar to the one observed in
the SBRM models \cite{lenz,jac,yan,klaus}. It also clearly shows that the
situation in our model is qualitatively different from the $\eta$-scaling in
the solid-state models with Anderson transition. There, in the limit of large
system size, only three
values $\eta=1$ (localized phase), $\eta=0$ (delocalized) and $\eta=\eta_A$
(at the transition) are possible \cite{shklo}. On the contrary in our case the 
scaling function varies smoothly from 1 to 0 with the rescaled 
transition matrix element $U/U_c$ for different system sizes $N$ which varied 
over more than two orders of magnitude. We relate this qualitative difference
between the two models to the fact that in the TBRIM all orbitals are coupled
by direct transitions whereas in the Anderson model, the hopping couples 
only nearby sites. Due to that the TBRIM is more similar to the 
SBRM models with broad band where many states are directly coupled.

\begin{figure}
\epsfxsize=3.7in
\epsfysize=2.6in
\epsffile{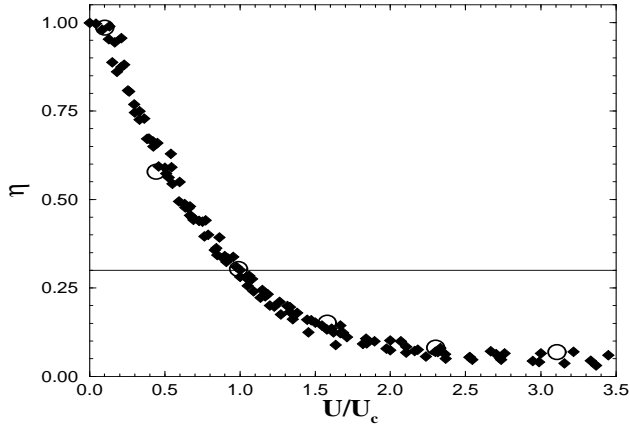}
\vglue 0.2cm
\caption{Dependence of $\eta$ on the rescaled interaction 
strength $U/U_c$ for
$2 \leq n \leq 8$, $4 \leq m \leq 80$, $1/40 \leq \nu 
\leq 1/2$ and $0.02 \leq U_c \leq 0.2$ (diamonds). Open circles show the
scaling close to the Fermi level (see text). The straight line marks
$\eta = \eta_c = 0.3$.} 
\label{fig2}
\end{figure}

The condition for the critical $U_c$ ($\eta_c=0.3$) allows to check the
theoretical prediction (\ref{uc}). The numerical data for which the number of
direct transitions varies over more than two
orders of magnitude are presented in Fig.
3. They give a clear confirmation of the estimate (\ref{uc}) giving
$C \approx 0.58$. The results of Figs. 1-3 show that for $U > U_c$ 
from (\ref{uc}) all nearby levels
are mixed by two-body interaction and $P(s)$ converges to the RMT result with
WD distribution. We stress that for large $m$ and $n$, the value of $U_c$ 
remains parametrically much larger than the multiparticle spacing $\Delta_n$.

\begin{figure}
\epsfxsize=3.7in
\epsfysize=2.6in
\epsffile{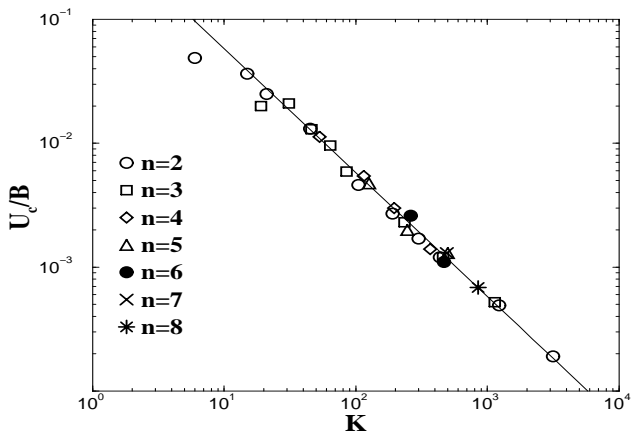}
\vglue 0.2cm
\caption{Dependence of the rescaled critical interaction strength $U_c/B$,
above which $P(s)$ becomes close to the Wigner-Dyson statistics,
on the number of directly coupled states $K$ for $4 \leq m \leq 80$
and $1/40 \leq \nu \leq 1/2$. The line shows the theory (1) with
$C=0.58$.
} 
\label{fig3}
\end{figure}

So far the results were obtained in the middle of the energy spectrum $E_h$
where all $K$ direct transitions are energetically allowed and effectively work. 
The situation becomes quite different close to the Fermi level. There, the
estimate (\ref{uc}) should be modified in the following way. First we should
take into account that the density of effectively coupled two-particle states 
$\rho_{2ef}$
becomes energy-dependent so that $\rho_{2ef}(\epsilon) \sim \epsilon/\Delta^2$
\cite{imry,tiq}. Secondly the number of effectively interacting particles is
also changed close to the Fermi level. Indeed as it is well known, at a 
temperature $T$, only $\delta n \sim T n/\epsilon_F \sim T/\Delta > 1$ particles 
interact near the Fermi surface. At this excitation energy 
$\epsilon \sim T < \epsilon_F$,
the density of two-particle states is $\rho_{2ef} \sim T/\Delta^2$. By
replacing in (\ref{uc}) $n$ by $\delta n$ and $\rho_2$ by $\rho_{2ef}$ we
obtain that at a given interaction strength the levels become mixed 
and $P(s)$ takes the WD form at a
temperature higher than the critical $T_{ch}$ given by 
\begin{equation}
\label{Tc}
T_{ch} \approx C_1 \Delta (\Delta/U)^{1/3}
\end{equation}
where $C_1$ is a numerical constant. The conditions of validity of this 
equation are $T_{ch} > \Delta \;\; (\delta n > 1) 
$ and $T_{ch} < \epsilon_F=n\Delta$
that corresponds to $n^{-3}  < U/\Delta < 1$.
It is also assumed that the WD statistics implies thermalization
with Fermi-Dirac statistics.
Such a conjecture looks quite natural, since the quantum chaos
should be related with excitation of many unperturbed modes and mixing.
Also without mixing of nearby levels and WD statistics the thermalization
is not possible since generally the Poisson distribution indicates an
existence of uncoupled parts in the whole system. As a result 
the thermalization does not exist below $T_{ch}$.

Since near the Fermi level
the total system energy counted from $E_g$
is $\delta E =E - E_g \approx T \delta n$, the relation
(\ref{Tc}) implies that the thermalization takes place only
for eigenstates with eigenenergies $E_{\lambda} = E_g + \delta E$
so that
\begin{equation}
\label{ec}
\delta E > \delta E_{ch} \approx C_1^2 \Delta (\Delta/U)^{2/3}
\end{equation}
The above restriction for $U$ requires $1 < \delta E/\Delta < n^2$.
This result shows that the $\eta$-parameter should depend on the
excitation energy. Indeed, our numerical data, extracted from
$P(s)$ computed in a small energy interval near a fixed
$\delta E$, clearly show that $\eta$ decreases with increasing 
excitation energy $\delta E$ (Fig. 4). Using the
relation (\ref{ec})  we can determine for a given $\delta E$ 
an effective $U_c$ value being $U_c = C_1^3\Delta (\Delta/\delta E)^{3/2}$.
The condition
$\eta(\delta E) = \eta_c = 0.3$ for the data of Fig. 4 at $n=6, m=12,
U/\Delta = 0.147$ gives $C_1 \approx 1.08$.
With the
value $C_1 = 1.08$ and the above dependence of $U_c$ on $\delta E$
we can check if the data of Fig. 4 will follow the general scaling
law of Fig. 2. For that in Fig. 2 we plot the $\eta$-values of Fig. 4
vs. the ratio $U/U_c$ with $U_c = 1.26 (\delta E)^{-3/2} $,
$C_1 = 1.08$ and $\Delta =1$ (open circles). The fact
that these data follow the scaling curve confirms the theoretical 
estimates (\ref{Tc}), (\ref{ec}) for the thermalization border.
The direct check of the dependence of $\delta E_{ch}$ on $U$ (insert in Fig.4)
also confirms the prediction (3).

\begin{figure}
\epsfxsize=3.7in
\epsfysize=2.6in
\epsffile{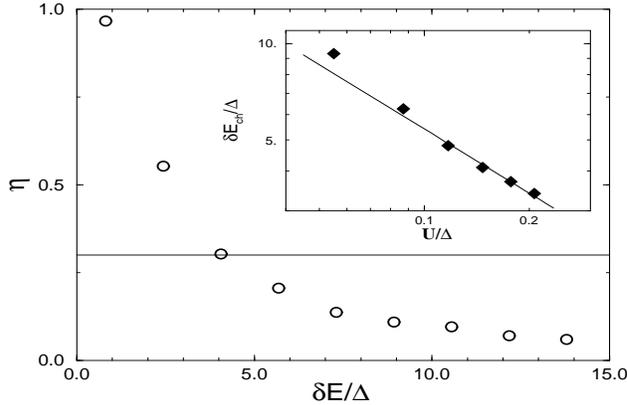}
\vglue 0.2cm
\caption{Dependence of $\eta$ on the rescaled excitation energy
$\delta E/ \Delta$ for $n=6, m=12$ and $ U/\Delta=0.147$ (o).
The straight line marks $\eta = \eta_c = 0.3$. Insert gives the 
numerically found dependence
of $\delta E_{ch}$ with $\eta=\eta_c=0.3$ on $U$ (diamonds), 
the straight line shows the theory (3) with $C_1=1.08$.
} 
\label{fig4}
\end{figure}

The obtained estimates for the quantum chaos border (\ref{Tc}), (\ref{ec})
can be applied to different finite interacting Fermi systems such
as complex nuclei with residual interaction,
atoms and molecules, clusters and quantum dots. Here we briefly discuss
the case of metallic quantum dots \cite{sivan}. In this case the interparticle
interaction is relatively weak so that $U/\Delta \sim 1/g$
with $g = E_c/\Delta \gg 1$ being the conductance of the dot and
$E_c$ the Thouless energy \cite{blanter}. According to (\ref{ec})
the thermalization will take place above the excitation energy
$\delta E_{ch} \sim \Delta g^{2/3}$. This is in a satisfactory agreement
with the experimental results \cite{sivan} where a dense spectrum
of excitations in dots with $g \sim 100$ appears at 
excitation energies $\delta E_{ch} \sim 10 \Delta$. We note that
our border for thermalization and chaos $\delta E_{ch}$ is higher than the
border for quasiparticle disintegration on many modes 
$\delta E_D \sim \Delta g^{1/2}$ proposed in \cite{alt,yan1}.
In our opinion the parametrically different dependence on $g$
suggested in \cite{alt,yan1} appears because the effect 
of energy redistribution between many excited modes was neglected
while the derivation of estimates (\ref{Tc}), (\ref{ec})
shows that it plays an important role. In addition, 
in the relations similar to (\ref{uc}) the authors
of \cite{alt,yan1} in fact used the first power of $n$, instead
of $n^2$, that according to our numerical data (Fig. 3) does not correspond to 
the regime with many excited modes.

In conclusion, our numerical results and analytical estimates
allowed to determine the border for emergence of quantum chaos 
and thermalization in finite interacting Fermi systems. 
Further investigation of this crossover in real systems is
highly desirable.


\end{document}